\documentclass[onecolumn]{article}%
\usepackage{amsmath}
\usepackage{amsfonts}
\usepackage{amssymb}
\usepackage{graphicx}
\usepackage{revsymb}%
\begin{document}

\title{A Note on the Asymptotic Limit of the Four Simplex}
\author{Suresh. K.\ Maran}
\maketitle

\begin{abstract}
Recently the asymptotic limit of the Barrett-Crane models has been studied by
Barrett and Steele. Here by a direct study, I\ show that we can extract the
bivectors which satisfy the essential Barrett-Crane constraints from the
asymptotic limit. Because of this the Schlaffi identity is implied by the
asymptotic limit, rather than to be imposed as a constraint.

\end{abstract}

The asymptotic limit \cite{AngMomClassLimit}, \cite{PonzanoReggeModel},
\cite{JWBRW} of the Barrett-Crane models \cite{BCReimmanion},
\cite{BCLorentzian} has been recently studied systematically in \cite{JWBCS}.
Here by a direct study, I\ show that we can extract the bivectors which
satisfy the essential Barrett-Crane constraints\footnote{Here by the essential
Barrett-Crane constraints \cite{BCReimmanion} I\ mean all the constraints
except those that relate to degeneracy conditions of a four simplex.} from the
asymptotic limit. Because of this the Schlaffi identity is implied by the
asymptotic limit, rather than to be imposed as a constraint as in
Ref:\cite{BCReimmanion}. Here I\ focus on the Riemannian Barrett-Crane model
\cite{BCReimmanion} only, but it can be generalized to the Lorentzian
Barrett-Crane model \cite{BCLorentzian}.

Consider the amplitude of a four-simplex\cite{BCReimmanion} with a real scale
parameter $\lambda$,%
\begin{align*}
Z_{\lambda} &  =(-1)^{\sum_{k<l}2J_{kl}}\int_{n_{k}\in S^{3}}%
{\displaystyle\prod\limits_{i<j}}
\dfrac{\sin(\lambda J_{ij}\theta_{ij})}{\sin(\theta_{ij})}%
{\displaystyle\prod\limits_{k}}
dn_{k},\\
&  =\frac{(-1)^{\sum_{k<l}2J_{kl}}}{(2i)^{10}}\int_{n_{k}\in S^{3}}%
{\displaystyle\prod\limits_{i<j}}
\sum\limits_{\varepsilon_{ij}=\pm1}\dfrac{\varepsilon_{ij}\exp(i\varepsilon
_{ij}\lambda(\rho J_{ij}\theta_{ij}))}{\sin(\theta_{ij})}%
{\displaystyle\prod\limits_{k}}
dn_{k},
\end{align*}
where the $\theta_{ij}$ is defined by $n_{i}.n_{j}=\cos(\theta_{ij})$. Here
the $\theta_{ij}$ is the angle between $n_{i}$ and $n_{j}$. The asymptotic
limit of $Z_{\lambda}(s)\ $under $\lambda\longrightarrow\infty$ is controlled by%

\begin{align*}
&  S(\{n_{i}\},\{J_{ij}\},\{q_{i}\})\\
&  =\sum_{i<j}\varepsilon_{ij}J_{ij}\theta_{ij}+\sum_{i}q_{i}(n_{i}.n_{i}-1),
\end{align*}
where the $q_{i}$ are the Lagrange multipliers to impose $n_{i}.n_{i}%
=1,\forall i$. My goal now is to find stationary points for this action. The
stationary value under the variation of $n_{j}$'s are determined by
\begin{subequations}
\begin{equation}
\sum_{~~~~~i\neq j}\varepsilon_{ij}J_{ij}\frac{\partial\zeta_{ij}}{\partial
n_{i}}+2q_{j}n_{j}=0,\forall j, \label{ext}%
\end{equation}
and $n_{j}.n_{j}=1,\forall j$ where the $j$ is a constant in the summation.%

\end{subequations}
\begin{equation}
\frac{\partial\zeta_{ij}}{\partial n_{i}}=\frac{n_{j}}{\sin(\zeta_{ij})},
\label{eq.dif.as}%
\end{equation}
where the vector index has been suppressed on both the sides.

Using equation (\ref{eq.dif.as}) in equation:(\ref{ext}) and taking the wedge
product of the equation with $n_{j}$ we have,
\[
\sum_{~~~~~i\neq j}\varepsilon_{ij}J_{ij}\frac{n_{i}\wedge n_{j}}{\sin
(\zeta_{ij})}=0,\forall j.
\]

If
\[
E_{ij}=\varepsilon_{ij}J_{ij}\frac{n_{i}\wedge n_{j}}{\sin(\zeta_{ij})},
\]
then the last equation can be simplified to%

\begin{equation}
\sum_{~~~~~i\neq j}E_{ij}=0,\forall j. \label{sumzero}%
\end{equation}

We now consider the properties of $E_{ij}:$

\begin{itemize}
\item Each $i$ represents a tetrahedron. There are ten $E_{ij}$'s, each one of
them is associated with one triangle of the four-simplex.

\item The square of $E_{ij}$:%
\begin{align*}
E_{ij}\cdot E_{ij}  &  =\frac{J_{ij}^{2}}{\sin^{2}(\zeta_{ij})}(n_{j}^{2}%
n_{i}^{2}-\left(  n_{i}\cdot n_{j}\right)  ^{2})\\
&  =\frac{J_{ij}^{2}}{\sin^{2}(\zeta_{ij})}(1-\left(  \cos(\zeta_{ij}\right)
^{2})\\
&  =J_{ij}^{2}.
\end{align*}

\item The wedge product of any two $E_{ij}$ is zero if they are equal to each
other or if their corresponding triangles belong to the same tetrahedron.

\item Sum of all the $E_{ij}$ belonging to the same tetrahedron are zero
according to equation (\ref{sumzero}).
\end{itemize}

It is clear that these properties contain the first four Barrett-Crane
constraints\cite{BCReimmanion}. So we have successfully extracted the
bivectors corresponding to the triangles of a general flat four-simplex in
Riemannian general relativity and the $n_{i}$ are the normal vectors of the
tetrahedra. The $J_{ij}$ are the areas of the triangle as one would expect.
Since we did not impose any non-degeneracy Barrett-Crane conditions
\cite{BCReimmanion}, it is not guaranteed that the tetrahedra or the
four-simplex have non-zero volumes.

The asymptotic limit of the partition function of the entire simplicial
manifold with triangulation $\Delta$ is%
\[
S(\Delta,\{n_{is}\in S^{3},J_{ij},\varepsilon_{ijs}\})=\sum_{i<j,s}%
\varepsilon_{ijs}J_{ij}\zeta_{ijs},
\]
where I\ have assumed variable $s$ represents the four simplices of $\Delta$
and $i,$ $j$ represents the tetrahedra. The $\varepsilon_{ijs}$ can be
interpreted as the orientation of the triangles. Each triangle has a
corresponding $J_{ij}$. The $n_{is}$ denote the vector associated with the
side of the tetrahedron $i$ facing the inside of a simplex $s$. Now there is
one bivector $E_{sij}$ associated with each side facing inside of a simplex
$s$ of a triangle $ij$ defined by%

\[
E_{ijs}=\varepsilon_{ijs}J_{ij}\frac{n_{js}\wedge n_{js}}{\sin(\zeta_{ijs})}.
\]
If the $n_{is}$ are chosen such that they satisfy stationary conditions%

\[
\sum_{~~~~~i\neq j}E_{ijs}=0,\forall j,s,
\]
and if
\[
\theta_{ij}=\sum_{s}\varepsilon_{ijs}\zeta_{ijs},
\]

then%

\begin{align*}
S(\Delta,\{J_{ij},\varepsilon_{ijs}\})  &  =\sum_{i<j,s}\varepsilon
_{ijs}J_{ij}\zeta_{ijs},\\
&  =\sum_{i<j}J_{ij}\theta_{ij}%
\end{align*}
can be considered to describe the Regge calculus \cite{ReggeCalc} for the
Riemannian general relativity. The angle $\theta_{ij}$ are the deficit angles
associated with the triangles and the $n_{is}$ are the vector normals
associated with the tetrahedra.

\end{document}